\documentclass[aps,pra,twocolumn,showpacs,preprintnumbers,amsmath,amssymb,superscriptaddress,longbibliography]{revtex4-2}
\usepackage{mathrsfs}
\usepackage{amsfonts}
\usepackage{amsmath}
\usepackage{amssymb}
\usepackage{natbib}
\usepackage{bbm}
\usepackage[normalem]{ulem}

\usepackage{xcolor}
\usepackage{graphicx}

\usepackage{hyperref}

\begin{document}
	\title{Interaction-Assisted Spin Coherence in a Multimode Trapped Atomic Ensemble}


	\author{Aleksandre Merabishvili}
	\affiliation{Department of Physics, College of William and Mary, Williamsburg, Virginia 23187, USA.}
	
	\author{Ebubechukwu O. Ilo-Okeke} \email{ebube@wm.edu}
	\affiliation{Department of Physics, College of William and Mary, Williamsburg, Virginia 23187, USA.}  

	\date{\today}
	
	\begin{abstract}	
		In a thermally occupied multimode trapped ensemble, mode-dependent detunings can strongly suppress Ramsey contrast even when the underlying dynamics remains coherent. We develop a weak-axis-resolved Hamiltonian theory in which collective mode spins experience longitudinal detunings and overlap-dependent exchange interactions. Without exchange, the distribution of detunings across thermally occupied modes produces coherent phase mixing and an algebraically decaying Ramsey-contrast envelope in the many-mode continuum regime. At the mean-field level, exchange generates a self-consistent restoring field that reduces phase dispersion and restores collective contrast. The crossover is governed by the competition between the collective exchange scale and the thermally induced detuning spread: when the late-time contrast is expressed in terms of their ratio, results at different temperatures collapse onto a common trend, and the crossover coupling follows the predicted inverse-square thermal scaling, with a numerical exponent of 2.06. Weak irreversible pure dephasing limits, but does not eliminate, this recovery, demonstrating that exchange refocuses coherent inhomogeneous phase mixing but cannot reverse irreversible phase noise.
	\end{abstract}

	\maketitle
	\section{Introduction}\label{sec:Introduction}
		
		Collective transverse-spin coherence is the resource that underlies Ramsey spectroscopy~\cite{ramsey1950,ramsey1985}, atomic clocks~\cite{barber2008,ludlow2015,ushijima2015,aeppli2024,zhang2024}, matter-wave interferometry~\cite{cronin2009,hammerer2010}, and many protocols in quantum sensing and quantum information \cite{ludlow2015,degen2017,pezze2018}. In a trapped atomic ensemble~\cite{garcia2006,deutsch2010,solaro2016}, this coherence is measured operationally through the collective transverse spin, or equivalently through the Ramsey-fringe contrast. Its preservation is therefore central to precision measurement and coherent collective-spin control. Interactions are often regarded as a source of frequency shifts~\cite{reeves2005,horikoshi2006} and decoherence~\cite{horikoshi2007,ilo-okeke2010}; however, in a multimode ensemble they can also reorganize the coherent spin dynamics and enhance the observable collective Ramsey contrast~\cite{deutsch2010,solaro2016}.
		
		A particularly important source of contrast loss in trapped gases is inhomogeneous phase evolution~\cite{reeves2005,horikoshi2006,stickney2007,kafle2011}. Atoms occupying different motional modes sample different spatial regions of the trap and can therefore experience different transition frequencies. The mode-resolved spins then accumulate different phases, causing the collective transverse spin to decrease even when the evolution within each mode remains fully coherent. This loss of contrast is therefore a phase-mixing process and is physically distinct from irreversible single-particle dephasing. Magnetic-field inhomogeneities, differential light shifts, and other position-dependent clock shifts can all generate such inhomogeneous broadening~\cite{deutsch2010,rosenbusch2009,kleinebuning2011,harber2002,widera2006}. Near a magic magnetic field, the linear sensitivity of the clock transition is suppressed, and the leading residual spatial shift may occur at higher order. When the residual field variation follows the trap potential, this produces a mode-dependent detuning across the thermally occupied motional states.
		
		The reversible character of this phase mixing is familiar from a conventional spin-echo sequence~\cite{hahn1950,carr1954,berman1997,eto2013,merkel2021,barnes2022,davis2023}. There, an externally applied control pulse reverses the relative phase winding generated by static or slowly varying detunings, allowing collective coherence to reappear at a later time. A decay of Ramsey contrast therefore need not imply an irreversible loss of microscopic coherence. Exchange-assisted self-rephasing shares this broad refocusing principle, but realizes it through a different mechanism: rather than imposing an externally timed reversal, exchange interactions continuously generate a state-dependent, self-consistent response that can reduce the phase dispersion among the mode-resolved spins. In this sense, exchange interactions can recover collective Ramsey contrast that has been suppressed by coherent phase mixing.
		
		This interaction-mediated recovery is commonly discussed as spin self-rephasing and has been observed and analyzed in a variety of trapped-atom settings \cite{deutsch2010,kleinebuning2011,maineult2012,solaro2016,bonnin2019,dupont-nivet2001}. Influential theoretical descriptions have formulated the dynamics through kinetic equations, coupled energy-class spin equations, and spin-wave approaches \cite{lhuillier1982,leggett1968,deutsch2010,kleinebuning2011,dupont-nivet2001}. These formulations provide effective descriptions of the collective spin motion and identify the exchange rate as the characteristic scale competing with inhomogeneous broadening.
		
		A complementary question is what mode-resolved many-body Hamiltonian underlies this restoring dynamics. Such a formulation places the inhomogeneous detuning and collisional exchange on the same microscopic footing. It identifies which contributions to contrast loss arise from coherent unitary evolution and which must be introduced separately as irreversible noise channels. It also connects the trap geometry, thermal mode occupations, and mode-overlap-dependent exchange couplings within a single model. Most importantly, the self-consistent exchange field appearing in the classical mode-spin dynamics can then be obtained as the mean-field limit of the underlying Heisenberg equations, rather than introduced as a phenomenological restoring term.
		
		In this work, we develop such a Hamiltonian description for a thermally occupied multimode trapped atomic ensemble. We retain the motional-mode dependence along the weakest trapping axis and incorporate the transverse degrees of freedom through thermal averaging of the interaction overlaps. Under the secular and forward-scattering approximations, the resulting model describes thermally weighted collective mode spins with different longitudinal detunings coupled by overlap-dependent isotropic exchange interactions. Near the magic magnetic field, the leading weak-axis contribution to the differential shift is quartic in position, producing a nonlinear dependence of the detuning on the axial mode index. This mode-resolved Hamiltonian provides a microscopic starting point for the exchange-assisted coherence dynamics and establishes a direct connection between the trapped-gas interactions and the collective restoring field.

			\begin{figure*}[t]
				\includegraphics[width = \textwidth]{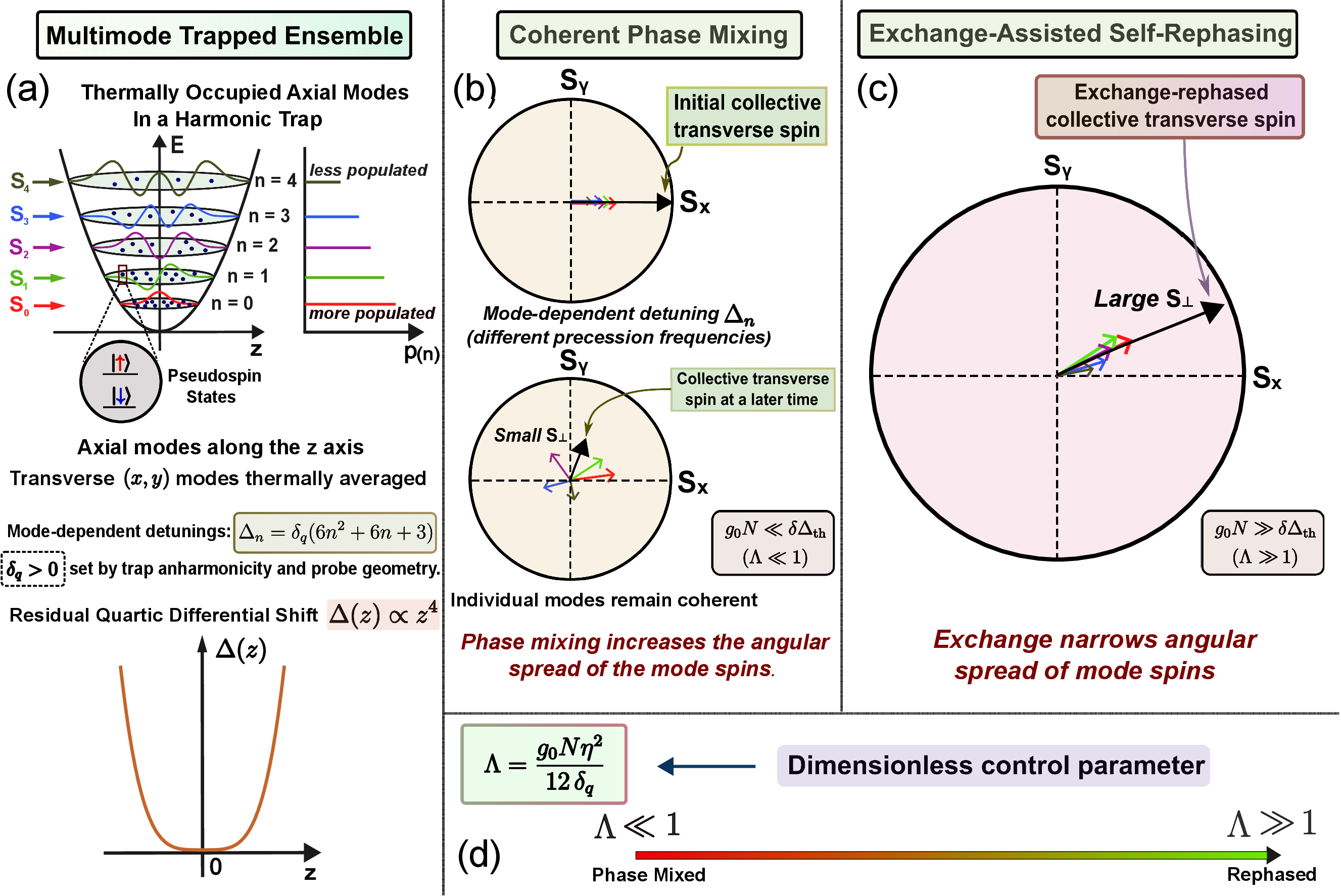}
				\caption{Conceptual picture of interaction-assisted spin coherence in a thermally occupied multimode trapped ensemble. (a) The weak-axis-resolved description retains the axial harmonic-oscillator modes explicitly, while the transverse $(x,y)$ degrees of freedom are incorporated through thermal averaging. Near the magic magnetic field, the residual differential shift is quartic in the axial coordinate, giving the occupied axial modes different detunings; each mode carries a collective pseudospin. (b) In the weak-exchange regime, initially aligned mode spins precess at different rates because of their mode-dependent detunings. The resulting coherent phase mixing increases their angular spread and reduces the collective transverse spin, while the individual mode spins remain coherent. (c) Exchange interactions generate a self-consistent restoring field that reduces the relative angular spread of the mode spins, thereby increasing the collective transverse spin and producing exchange-assisted self-rephasing. (d) The crossover from the phase-mixed to the rephased regime is organized by the dimensionless exchange-to-inhomogeneity ratio $\Lambda=g_0N\eta^2/(12\delta_q)$, with $\Lambda\ll1$ corresponding to weak exchange and $\Lambda\gtrsim1$ marking the onset of substantial interaction-assisted contrast recovery.}\label{fig1}
			\end{figure*}
		
		Figure~\ref{fig1} summarizes the physical picture and scaling structure developed in this work. Thermal occupation of axial motional modes produces a distribution of mode-dependent detunings and hence coherent phase mixing of the corresponding collective spins. Exchange interactions generate a self-consistent restoring response that reduces this phase dispersion and enhances the collective transverse spin. The crossover between these regimes is governed by the competition between the collective exchange scale and the thermally induced detuning spread. This competition defines the dimensionless exchange-to-inhomogeneity ratio \(\Lambda\), which serves as the organizing parameter for the interaction-assisted coherence dynamics analyzed below.

        The remainder of the paper is organized as follows. In Sec.~\ref{sec:PhysicalSystem}, we derive the weak-axis-resolved Hamiltonian for a thermally occupied trapped ensemble, retaining the mode-dependent detuning along the axial direction and the forward-scattering exchange interaction between collective mode spins. In Sec.~\ref{sec:ThermalDephasing}, we establish the noninteracting phase-mixing baseline and show that, in the many-mode continuum regime, the Ramsey contrast develops an algebraically decaying stationary-phase envelope rather than an exponential decay. In Sec.~\ref{sec:AssistedRephasing}, we derive the mean-field mode-spin dynamics and show that exchange produces a self-consistent restoring torque. We then demonstrate that the exchange-to-inhomogeneity ratio organizes the crossover and reveals a scaling collapse across temperatures, while the operational crossover coupling follows the thermal scaling predicted from the inhomogeneous broadening. In Sec.~\ref{sec:Robustness}, we include irreversible pure dephasing and distinguish exchange-assisted refocusing of coherent phase mixing from irreversible loss of transverse coherence.

	\section{Physical System and Model \label{sec:PhysicalSystem}}
		
		Consider a gas containing $N$ neutral atoms with two internal pseudospin states $|\uparrow\rangle,\,|\downarrow\rangle$, confined in a harmonic trap, $V_\mathrm{trap}(\mathbf{r})$. The reduced weak-axis description used below is illustrated schematically in Fig.~\ref{fig1}. The energy of the $n${th} mode in the trap is $E_n$ with a corresponding state $\phi_n(\mathbf{r})$. Let the operators $\hat{c}^\dagger_{\uparrow n}$ and $\hat{c}^\dagger_{\downarrow n}$ create and  $\hat{c}_{\uparrow n}$, $\hat{c}_{\downarrow n}$ destroy atoms in the internal states $\uparrow$ and $\downarrow$ in mode $n$. The total occupation of mode $n$ is $\hat{N}_n = \hat{c}^\dagger_{\uparrow n}\hat{c}_{\uparrow n} + \hat{c}^\dagger_{\downarrow n}\hat{c}_{\downarrow n}$. We define the following collective spin operators for the mode
		\begin{align}
			\label{eq:SpinZ}
			S^z_n &=  \frac{\hat{c}^\dagger_{\uparrow n}\hat{c}_{\uparrow n} - \hat{c}^\dagger_{\downarrow n}\hat{c}_{\downarrow n}}{2},\\
			\label{eq:SpinPlus}
			S^{+}_n &= \hat{c}^\dagger_{\uparrow n}\hat{c}_{\downarrow n}, \\
			\label{eq:SpinMinus}
			S^{-}_n &= \hat{c}^\dagger_{\downarrow n}\hat{c}_{\uparrow n} .
		\end{align}
		These operators obey the spin commutation relations
		\begin{equation}
			\label{eq:CommunationRelations}
			\left[S^z_n,S^\pm_n \right] = \pm S^{\pm}_n,\qquad \left[S^+_n,S^-_n \right] = 2S^z_n. 
		\end{equation}
		
		The two internal clock states generally experience different Zeeman shifts in the presence of a magnetic field. Consequently, spatial variations of the magnetic field produce a position-dependent transition frequency across the trapped ensemble.

	\subsection{Weak-Axis-Resolved Model}
		We now introduce a reduced description that retains the essential competition between inhomogeneous phase mixing and collisional spin exchange. The full trapped gas is three-dimensional, but we resolve the dominant mode dependence of the detuning along the weakest trapping axis, taken to be the $z$ direction. The transverse degrees of freedom are incorporated through thermal averaging of the interaction overlap, while the axial modes remain explicit. The resulting weak-axis-resolved model assigns a collective spin to each axial mode, with each mode subject to a mode-dependent detuning and an exchange field generated by the other occupied modes.
		
		\subsubsection{Quartic inhomogeneous detuning}
			
			The derivation of the reduced inhomogeneous Hamiltonian in Eq.~\eqref{eq:InhomogeneousHamil} is given in Appendix~\ref{sec:InhomogeneousDetuning}. Here we summarize the physical origin of the mode-dependent detuning and specify the weak-axis contribution used in the main text.
	
			Near the magic magnetic field $B_m$, the linear sensitivity of the clock transition to magnetic-field fluctuations vanishes. Expanding the transition frequency about $B_m$, the leading dependence is therefore quadratic in the magnetic-field deviation. If the field at the trap center is tuned close to $B_m$, the residual field variation near the trap minimum is quadratic in position, so that the leading spatially dependent clock shift is quartic in position. As a minimal reduced model, we isolate the contribution from the weakest trapping axis. Writing the axial harmonic potential as
			\begin{equation}
				V_z(z)=\alpha_z z^2, \qquad \alpha_z=\frac{1}{2}M\omega_z^2,
			\end{equation}
			we approximate the differential shift by the leading quartic contribution
			\begin{equation}
				\Delta(z) = \kappa V_z^2(z) = \kappa \alpha_z^2 z^4 .
			\end{equation}
			Here $\alpha_z$ denotes the axial trap curvature, and $\kappa$ is the quartic differential-shift coefficient that converts the squared trap potential into a position-dependent detuning.
	
			Within a secular, or random-phase, approximation, the single-particle Hamiltonian becomes [see \eqref{eq:Dephasing}] 
			\begin{equation}
				\label{eq:InhomogeneousHamil}
				\hat H^\Delta = \hbar \sum_n \tilde{\Delta}_n S_n^z ,
			\end{equation}
			where the mode-dependent shift $\tilde{\Delta}_n\equiv\tilde{\Delta}_{nn}$ is obtained by projecting $\Delta(z)$ onto axial harmonic-oscillator mode $n$:
			\begin{equation}
				\label{eq:DeltaNCompact}
				\tilde{\Delta}_n = \int dz\, |\phi_n(z)|^2\Delta(z) = \tilde{\delta}_q \left( 6n^2+6n+3 \right),
			\end{equation}
			with
			\begin{equation}
				\label{eq:DephasingStrength}
				\tilde{\delta}_q = \kappa\alpha_z^2 \left( \frac{\hbar}{2M\omega_z} \right)^2 .
			\end{equation}
			The mode-independent part of $\tilde{\Delta}_n$ produces a common frequency shift, whereas the $n$-dependent part produces the inhomogeneous phase evolution across the occupied axial modes.

		\subsubsection{Forward-scattering interaction and transverse averaging}
			We next construct the interaction Hamiltonian used in the reduced model. The full derivation, starting from the contact interaction and projecting onto trap modes, is given in Appendix~\ref{sec:ManyBody}. Here we state the forward-scattering form retained in the main text and describe the thermal averaging used to incorporate the transverse modes.
	
			Atoms in the trap also experience contact two-body interactions. The collision strength between atoms in internal states $\sigma$ and $\sigma'$ is characterized by $g_{\sigma\sigma'} =4\pi\hbar^2a_{\sigma\sigma'}/M$, 	where $a_{\sigma\sigma'}$ is the corresponding $s$-wave scattering length~\cite{vankempen2002,chin2010}. In the following, we assume isotropic interactions, $a_{\uparrow\uparrow} = a_{\downarrow\downarrow} = a_{\uparrow\downarrow}$. Within the secular or random-phase approximation, we retain only collision processes that preserve the set of occupied motional modes, namely the direct and exchange channels. The resulting effective spin Hamiltonian, as given in~\eqref{eq:InterSpinExchangeHamiltonian}, is
			\begin{equation}
				\label{eq:EffectiveSpinHamiltonian}
				\hat H^{\rm int} = g_{\rm ex}U^\perp \sum_{n<m} U^z_{nm} \left[ S_n^zS_m^z + \frac{1}{2} \left( S_n^+S_m^- 	+ S_n^-S_m^+ \right) \right], 
			\end{equation}
			where $g_{\rm ex}=2g_{\uparrow\downarrow}$. For a separable harmonic trap, the spatial overlap factorizes into transverse and axial parts,
			\begin{align}
				\label{eq:OverlapIntPerp}
				U^\perp_{\nu\mu} &=  \int dx\,dy\, \left|\phi^\perp_\nu(x,y)\right|^2 \left|\phi^\perp_\mu(x,y)\right|^2, \\
				\label{eq:OverlapIntZ}
				U^z_{nm} &= \int dz\, \left|\phi^z_n(z)\right|^2 \left|\phi^z_m(z)\right|^2 .
			\end{align}
			Here $\nu=(n_x,n_y)$ and $\mu=(m_x,m_y)$ label transverse modes, while $n,m$ label axial modes. The reduced model retains the axial mode dependence of $U^z_{nm}$ while incorporating the transverse degrees of freedom through the thermal average
			\begin{equation}
				\label{eq:EffectiveOverlap}
				U^\perp = \frac{ \sum_{\nu,\mu} p^\perp_\nu p^\perp_\mu U^\perp_{\nu\mu} }{ \sum_{\nu,\mu} p^\perp_\nu p^\perp_\mu }.
			\end{equation}
			The normalized transverse thermal weight is
			\begin{equation}
				p^\perp_\nu = \frac{e^{-\beta E^\perp_\nu}}{Z_\perp}, \qquad Z_\perp = \sum_\nu e^{-\beta E^\perp_\nu},
			\end{equation}
			with
			\begin{equation}
				E^\perp_\nu = \hbar\omega_x \left( n_x+\frac{1}{2} \right) + \hbar\omega_y \left( n_y+\frac{1}{2} 	\right),
			\end{equation}
			and $\beta=1/k_BT$.

		\subsubsection{Dimensionless reduced Hamiltonian}
	
			We now introduce the dimensionless quantities used in the final reduced Hamiltonian. The dimensionless mode-dependent detuning is
			\begin{equation}
				\label{eq:DimensionlessFreqDef}
				\Delta_n = \frac{\tilde{\Delta}_n}{\omega_z} = \delta_q \left( 6n^2+6n+3 \right),
			\end{equation}
			where $\delta_q = \tilde{\delta}_q/\omega_z$. The dimensionless inverse-temperature parameters are
			\begin{equation}
				\eta_i = \frac{\omega_i}{\omega_z}\eta, \qquad \eta = \beta\hbar\omega_z
			\end{equation}
			for $i=x,y$. The residual axial mode dependence of the interaction is contained in the normalized overlap
			\begin{equation}
				\label{eq:DimensionlessOverlapDef}
				\tilde U^z_{nm} = \frac{ U^z_{nm}(\eta) }{ \overline U^z(\eta) },
			\end{equation}
			where
			\begin{equation}
				\label{eq:ThermalAverageAxialOverlap}
				\overline U^z(\eta) = \frac{ \sum_{a<b} p_a(\eta)p_b(\eta) U^z_{ab}(\eta) }{ \sum_{a<b} 		p_a(\eta)p_b(\eta) }.
			\end{equation}
			With this convention, the effective dimensionless interaction strength is
			\begin{equation}
				\label{eq:g0Def}
				g_0 = \frac{ g_{\rm ex} U^\perp(\eta_x,\eta_y) 	\overline U^z(\eta) }{ \hbar\omega_z }.
			\end{equation}
			Thus $g_0$ sets the thermally averaged exchange scale, while $\tilde U^z_{nm}$ captures the residual axial channel-to-channel variation about this average.
	
			Using Eqs.~\eqref{eq:InhomogeneousHamil} and \eqref{eq:EffectiveSpinHamiltonian}, the effective dimensionless many-body Hamiltonian, $\mathcal{H} = \left( H^\Delta + H^{\rm int} \right) /\hbar\omega_z$, is
			\begin{equation}
				\label{eq:DimensionlessHamil}
				\begin{split}
					\mathcal{H}  = 	&\sum_n \Delta_n S_n^z 	+ g_0 \sum_{n<m} \tilde U^z_{nm} \Big[ S_n^zS_m^z \\
					&\qquad\qquad + \frac{1}{2} \left( S_n^+S_m^- + S_n^-S_m^+ \right) \Big].
				\end{split}
			\end{equation}
			This Hamiltonian provides the reduced microscopic description used in the remainder of the paper: the first term generates mode-dependent coherent phase evolution, while the second couples the mode spins through overlap-dependent exchange interactions.

	\section{Noninteracting Thermal Dephasing}\label{sec:ThermalDephasing}
		We first establish the noninteracting baseline for the collective spin dynamics. In this limit, the loss of Ramsey contrast arises entirely from coherent phase mixing among motional modes with different detunings.  
		
		The initial state is taken to be a product of spin coherent states, with the atoms in each occupied motional mode prepared along the positive $x$-axis of the Bloch sphere. For a fixed occupation configuration $\{N_n\}$, the initial state is
		\begin{equation}
			\lvert\chi_{\rm sp};\{N_n\}\rangle 	= \bigotimes_n \lvert\chi_{\rm sp};N_n\rangle ,
		\end{equation}
		where $\lvert\chi_{\rm sp};N_n\rangle$ denotes a collective spin coherent state of $N_n$ atoms in mode $n$ pointing along the $x$-axis,
		\begin{equation}
			\label{eq:CoherentState}
			\lvert\chi_{\rm sp};N_n\rangle =\sum_{m=0}^{N_n} \sqrt{\frac{N_n!}{2^{N_n}m!(N_n-m)!} }  \lvert m, N_n-m\rangle
		\end{equation} 
		Including the statistical distribution over motional occupations, the initial density matrix is
		\begin{equation}
			\rho_0 	= \sum_{\{N_n\}} P(\{N_n\}) \bigotimes_n \left[ |\chi_{\rm sp};N_n\rangle \langle \chi_{\rm sp};N_n| \right].
		\end{equation}

		In the absence of interactions, $g_\mathrm{0} =0$ in ~\eqref{eq:DimensionlessHamil}, and the system evolves under the inhomogeneous dephasing Hamiltonian 	
		\begin{equation}
			\label{eq:DephasingHamiltonian}
			\mathcal{H}^\Delta = \sum_n \Delta_n S_n^z .
		\end{equation}
		Because the terms in $\mathcal{H}^\Delta$ commute, the evolution factorizes exactly over the motional modes. Introducing the dimensionless time $\tau =\omega_z t$, the state at time $\tau$ is
		\begin{equation}
			\label{eq:StateEvol}
			\begin{split}
				\rho(\tau) =& \sum_{\{N_n\}} P(\{N_n\}) \\
				&\bigotimes_n \left[ e^{-i\Delta_n\tau S_n^z} \lvert \chi_{\rm sp};N_n\rangle \langle \chi_{\rm sp};N_n\rvert e^{i\Delta_n\tau S_n^z} \right].
			\end{split}
		\end{equation}
		Each motional mode therefore accumulates a phase at a rate set by its detuning $\Delta_n$. The resulting reduction of the collective transverse spin is a coherent phase-mixing effect caused by the distribution of mode-dependent detunings.
		
		At the end of the interferometric sequence, a $\pi/2$ pulse about the $y$-axis is applied. Defining
		\begin{equation}
			S^y=\sum_n S_n^y,
		\end{equation}
		the final state is
		\begin{equation}
			\rho_f =  e^{-i\frac{\pi}{2}S^y} \rho(\tau) e^{i\frac{\pi}{2}S^y}.
		\end{equation}
		Since the pulse factorizes over modes, this can equivalently be written as
		\begin{equation}
			\label{eq:FinalState}
			\begin{split}
				\rho_f = & \sum_{\{N_n\}} P(\{N_n\}) \bigotimes_n \Big[	e^{-i\frac{\pi}{2}S_n^y} e^{-i\Delta_n\tau S_n^z}\\
				&   \lvert\chi_{\rm sp};N_n\rangle \langle \chi_{\rm sp};N_n\rvert e^{i\Delta_n\tau S_n^z} e^{i\frac{\pi}{2}S_n^y} \Big].
			\end{split}
		\end{equation}

		The measured collective spin components are
		\begin{equation}
			\label{eq:CollectiveAverages}
			\langle S^\alpha\rangle = \mathrm{Tr}(\rho_f S^\alpha),  \qquad S^\alpha=\sum_n S_n^\alpha,
		\end{equation}
		where $\alpha=x,y,z$. 
%
		For the initial spin coherent state polarized along $x$, one obtains
		\begin{equation}
			\label{eq:ExpectationValues}
			\begin{split}
				\langle S^x\rangle &= 0, \\
				\langle S^y\rangle &= \sum_n \sum_{\{N_n\}} P(\{N_n\}) \frac{N_n}{2} \sin\left(\Delta_n\tau\right), \\
				\langle S^z\rangle &= -\sum_n \sum_{\{N_n\}} P(\{N_n\}) \frac{N_n}{2} \cos\left(\Delta_n\tau\right).
			\end{split}
		\end{equation}

		For independent thermal occupations, the ensemble average over $N_n$ replaces $N_n$ by its mean value $\bar{N}_n=Np_n$, $\sum_{N_n} P(N_n)\,N_n = \bar{N}_n=Np_n $. Using the Boltzmann distribution
		\begin{equation}
			\label{eq:BoltzDistri}
			p_n = (1-e^{-\eta})e^{-n\eta},
		\end{equation}
		the nonzero spin averages become
		\begin{equation}
			\label{eq:NonVanishingExpectations}
			\begin{split}
				\langle S^y\rangle &=  \frac{N}{2} (1-e^{-\eta}) \sum_{n=0}^{\infty} e^{-n\eta} \sin\left(\Delta_n\tau\right), \\
				\langle S^z\rangle &= -\frac{N}{2} (1-e^{-\eta}) \sum_{n=0}^{\infty} e^{-n\eta} 		\cos\left(\Delta_n\tau\right).
			\end{split}
		\end{equation}
		These expressions make explicit that the collective spin dynamics is governed by the thermal distribution over motional modes. At low temperature, $\eta\gtrsim 1$, only a few modes are occupied and the discreteness of the sum is essential. In the opposite regime, $\eta\ll1$, many modes contribute, with a characteristic thermal mode number
		\begin{equation}
			\label{eq:ThermalScale}
			n_{\rm th}\sim \frac{1}{\eta}.
		\end{equation}
		
		For the quartic detuning~\eqref{eq:DimensionlessFreqDef}, $\Delta_n = \delta_q(6n^2+6n+3)$, the mode-dependent part of the phase is $ \phi(n) = 6\delta_q\tau n(n+1)$. The phase variation between neighboring thermally occupied modes is therefore approximately $\phi'(n) \simeq	12\delta_q\tau n $. Evaluating this at the characteristic thermal mode number $n\sim n_{\rm th}\sim1/\eta$, the collective spin begins to lose phase coherence when
		\begin{equation}
			\frac{12\delta_q\tau}{\eta}\sim 1.
		\end{equation}
		This gives the characteristic collapse time
		\begin{equation}
			\label{eq:CollapseTime}
			\tau_{\rm c} \sim \frac{\eta}{12\delta_q}.
		\end{equation}
		This estimate shows that the inhomogeneous phase-mixing time decreases as the thermal mode distribution broadens. 
		
		
		On the other hand, given that the local slope of the detuning $\Delta_n$ near the thermal modes is $\Delta'_n \sim 12\delta_q /\eta$, the corresponding detuning spread across the occupied modes $\delta n \sim 1/\eta$ is  $\delta \Delta_\mathrm{th} = \Delta'_n\, \delta n$,
		\begin{equation}
			\label{eq:DetuningSpread}
			\delta \Delta_\mathrm{th} \sim \frac{12\delta_q}{\eta^2}.
		\end{equation}
		Thus, the phase spread $\delta \Delta_\mathrm{th} \tau_\mathrm{mix}$ across the thermally occupied range becomes of order unity $\delta \Delta_\mathrm{th} \tau_\mathrm{mix} \sim 1$ on a time scale  
		\begin{equation}
			\label{eq:PhaseMixingTime}
			\tau_\mathrm{mix} \sim \frac{\eta^2}{12\delta_q }.
		\end{equation}
%
		This scale marks the onset of order-one phase variation across the thermally occupied detuning bandwidth, but it precedes the observable collapse of the ensemble contrast. At $\tau_{\mathrm{mix}}$, the phase difference between neighboring modes near the characteristic thermal occupation remains only of order $\eta\ll1$. Since \begin{equation}
			\label{eq:CollapseMixingRelation}
			\tau_\mathrm{mix} = \eta \tau_c,
		\end{equation} 
		the high-temperature regime admits a broad interval in which phase mixing is already appreciable while the collective Ramsey contrast remains substantial.
		
		In the high-temperature regime $\eta\ll 1$, the summand~\eqref{eq:NonVanishingExpectations} varies slowly over the thermally occupied modes, and the discrete sum can be approximated by an integral. This gives the asymptotic expressions
		\begin{align}
			\label{eq:SyApproxSimplified}
			\langle S^y\rangle &= \frac{N}{2} \mathcal C_\mathrm{asy}(\tau) \sin\left[ \left(  \frac{3}{2}\delta_q + 	\frac{\tau_\mathrm{mix}}{2\tau^2} \right)\tau + \frac{\pi}{4} \right], \\
			\label{eq:SzApproxSimplified}
			\langle S^z\rangle &= -\frac{N}{2} \mathcal C_\mathrm{asy}(\tau) \cos\left[ \left(  \frac{3}{2}\delta_q + 	\frac{\tau_\mathrm{mix}}{2\tau^2} \right)\tau + \frac{\pi}{4} \right].
		\end{align}
		The normalized contrast is
		\begin{equation}
			\label{eq:Contrastdefinition}
			\mathcal C(\tau) = \frac{2}{N} \sqrt{ \langle S^y\rangle^2 + \langle S^z\rangle^2 }.
		\end{equation}
		
			\begin{figure}[!t]
				\includegraphics[width = \columnwidth]{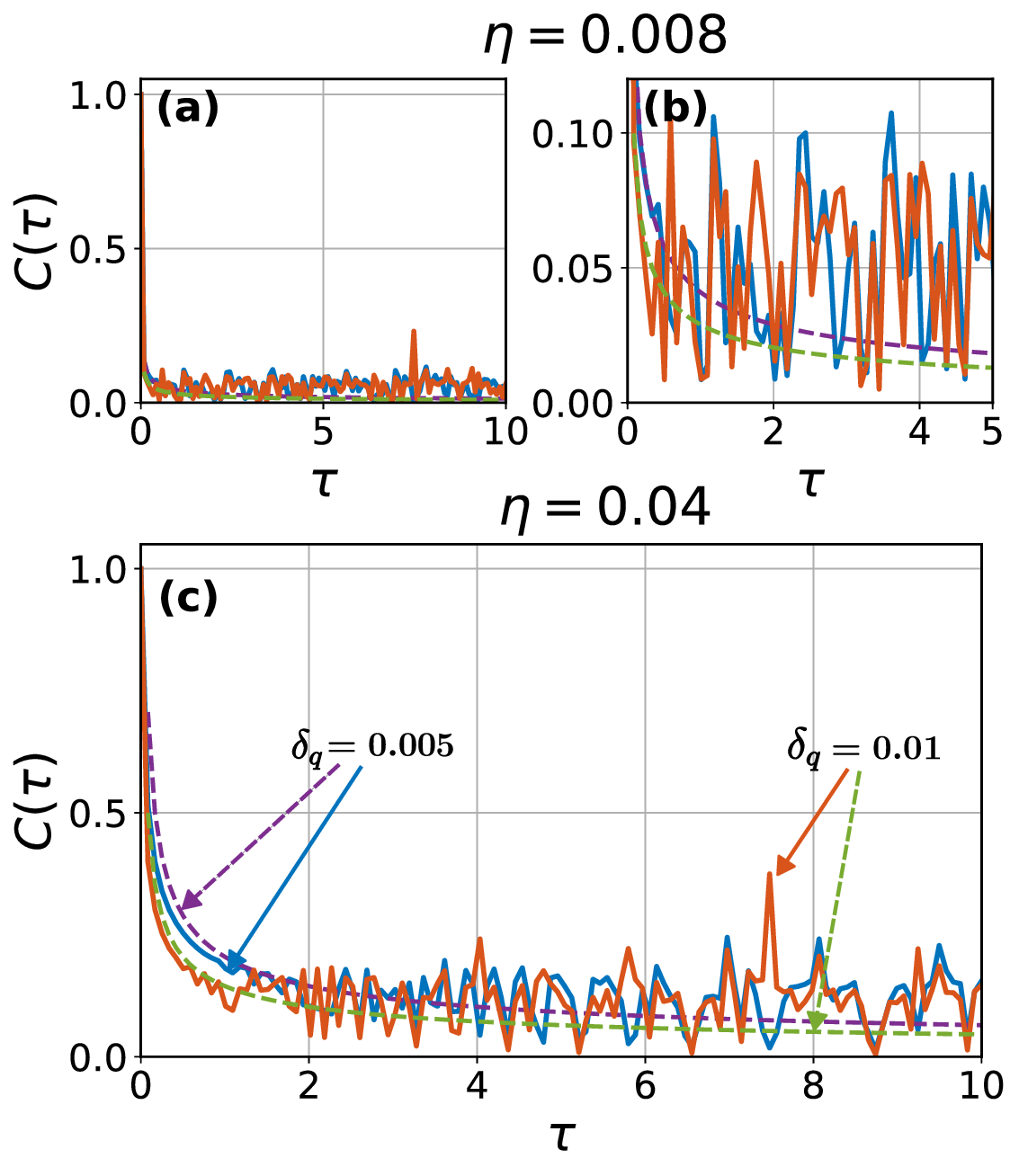}
				\caption{Comparison between the discrete-mode numerical contrast and the continuum analytic asymptote.
					Solid curves show the numerically evaluated contrast $C(\tau)$~\eqref{eq:Contrastdefinition}, while dashed curves show the analytic approximation~\eqref{eq:AsymptoticContrast}, plotted only in its asymptotic regime \(\tau\gtrsim 2\tau_\mathrm{mix}\). Panel (a) shows the small-$\eta$ high-temperature regime, and (b) zooms into the low-contrast regime of (a). Panel (c) shows the intermediate temperature regime. All parameters are as shown in the figure.
				}\label{fig2}
			\end{figure}
		The contrast results are shown in Fig.~\ref{fig2}.  The solid curves are obtained from the discrete mode sum, while the dashed curves show the continuum stationary-phase approximation.  The approximation is controlled in the small-$\eta$ regime,
		\begin{equation}
			0<\eta\ll 1,
		\end{equation}
		where many motional modes contribute to the thermal average and the mode sum may be replaced by a continuum integral.  In this regime, inhomogeneous broadening characterized by \(\delta_q\) produces algebraic phase mixing rather than exponential decay.  The late-time contrast takes the asymptotic form
		\begin{equation}
			\label{eq:AsymptoticContrast}
			\mathcal{C}_{\rm asy}(\tau) = 	\sqrt{\frac{\pi}{6\delta_q \tau}}\sinh\left(\frac{\eta}{2}\right),
		\end{equation}
		so that
		\begin{equation}
			\mathcal C_{\rm asy}(\tau)\propto \tau^{-1/2}.
		\end{equation}
		
		Figure~\ref{fig2}(a) shows the small-$\eta$, high-temperature regime, where the continuum approximation captures the algebraic envelope of the numerical contrast after the initial transient.  Figure~\ref{fig2}(c) shows an intermediate value of $\eta$, for which the same asymptotic trend remains visible, although deviations from the continuum curve become more pronounced.  These deviations arise because the numerical result retains discrete finite-mode interference that is smoothed out in the continuum approximation.  Figure~\ref{fig2}(b) magnifies the low-contrast regime of panel (a), showing that the asymptotic expression describes the slowly decaying envelope rather than the residual finite-mode fluctuations.

		Beyond the algebraic suppression of contrast, the late-time asymptotic form also contains a slowly varying phase. The stationary-phase approximation gives a phase
		\begin{equation}
			\Phi(\tau)=\frac{3}{2}\delta_q\tau+\frac{\tau_\mathrm{mix}}{2\tau}+\frac{\pi}{4},
		\end{equation}		
		so that the instantaneous frequency approaches
		\begin{equation}			
			\omega_{\rm inst}(\tau)=\frac{d\Phi}{d\tau} =\frac{3}{2}\delta_q-\frac{\tau_\mathrm{mix}}{2\tau^2}.
		\end{equation}
		Thus the asymptotic dynamics approaches an effective frequency
		\begin{equation}
			\omega_{\rm eff}=\frac{3}{2}\delta_q,
		\end{equation}
		with a transient correction that decays as $1/\tau^2$.  The universal phase offset $\pi/4$ reflects the stationary-phase origin of the late-time dynamics. Over the time window shown in Fig.~\ref{fig2}, this phase evolution is slow compared with the residual discrete-mode interference, explaining the small fluctuations of the numerical contrast about the smooth asymptotic envelope.
		
		\begin{figure}[!t]
			\includegraphics[width = \columnwidth]{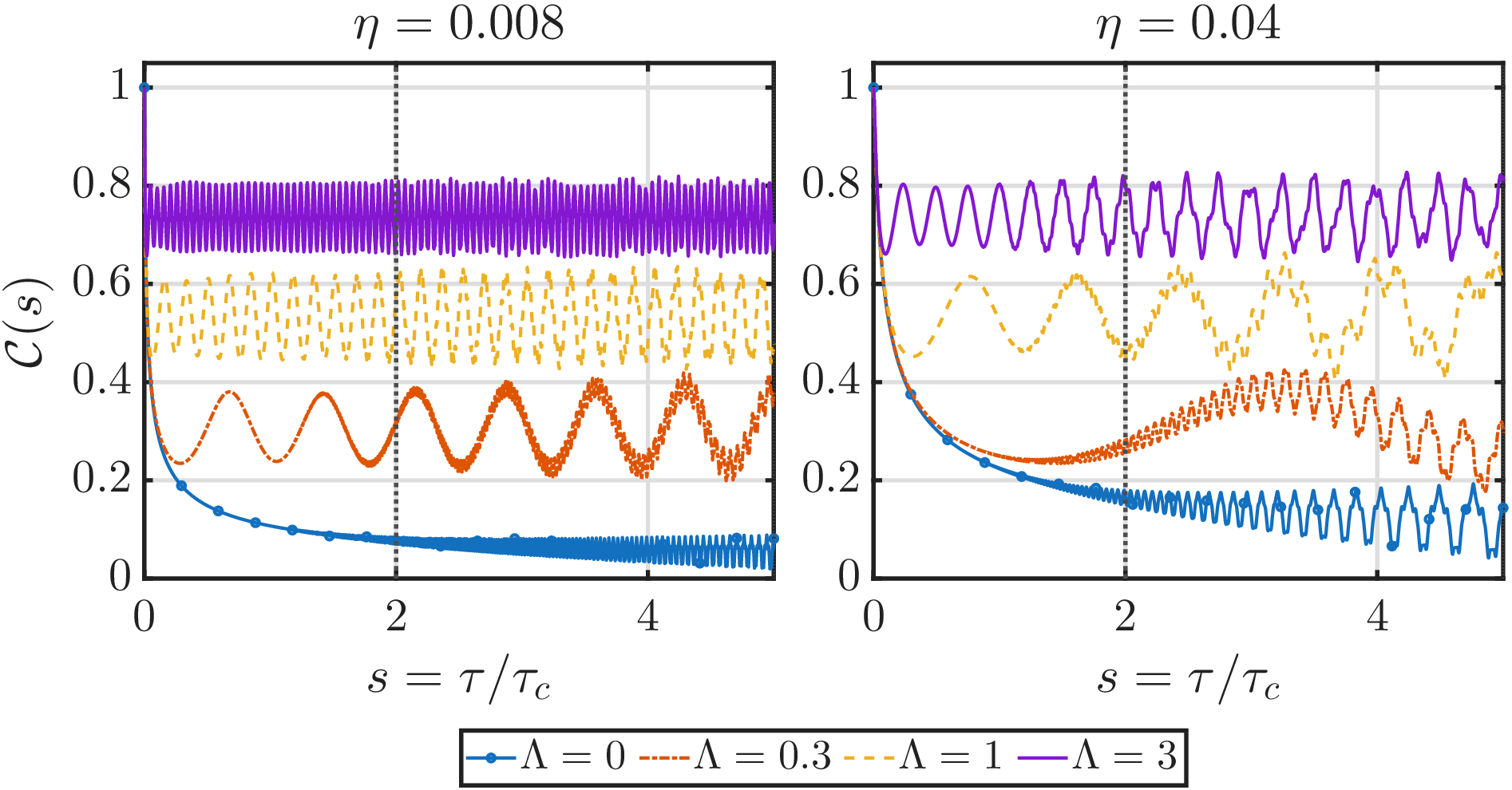}
			\caption{Representative Ramsey-contrast dynamics for two temperatures. The Ramsey contrast $\mathcal C(s)$, with $s=\tau/\tau_c$, is shown for $\eta=0.008$ and $\eta=0.04$ at several values of $\Lambda$~\eqref{eq:ExchangeDephasingParameter}. The dotted vertical line at $s=2$ marks the beginning of the late-time averaging window used to define $\bar{\mathcal C}_{\rm late}$.
			}\label{fig3}
		\end{figure}

	\section{Exchange-Assisted Self-Rephasing}\label{sec:AssistedRephasing}
		For large atom number and many thermally occupied motional modes, the exact many-body dynamics generates a hierarchy of spin correlations that is computationally expensive to retain. We therefore consider the mean-field limit, in which each occupied mode is represented by a classical collective spin. This approximation is motivated by the experimentally relevant regime of large atom number, for which relative collective fluctuations scale as $1/\sqrt{N}$. More specifically, the mean-field description is expected to be most reliable for the thermally important modes with large occupations $N_k$.
		
		We define the normalized mode-resolved spin variables
		\begin{equation}
			u_k^\alpha = \frac{\langle S_k^\alpha\rangle}{N_k}, \qquad N_k=Np_k,
		\end{equation}
		where $p_k$ is the thermal weight of mode $k$,~\eqref{eq:BoltzDistri}. With this convention, an initially $x$-polarized spin coherent state has $u_k^+=1/2$ and $u_k^z=0$. The mean-field approximation amounts to factorizing inter-mode spin correlations as
		\begin{equation}
			\langle S_k^\alpha S_n^\beta\rangle \approx \langle S_k^\alpha\rangle \langle S_n^\beta\rangle , \qquad k\neq n .
		\end{equation}
		Applying this factorization to the Heisenberg equations of motion using~\eqref{eq:DimensionlessHamil} gives the classical mode-spin equations
		\begin{align}
			\label{eq:SplusEvolModified}
			\frac{du_k^+}{d\tau} & = i\Delta_k u_k^+ + ig_{0}N \sum_{n\neq k} \tilde{U}^z_{nk}p_n \left( u_k^+u_n^z - u_k^z u_n^+ 	\right),\\
			\label{eq:SzEvolModified}
			\frac{du_k^z}{d\tau}
			&=  \frac{ig_{0}N}{2} \sum_{n\neq k} \tilde{U}^z_{nk} p_n \left( u_n^+u_k^- - u_n^-u_k^+ \right).
		\end{align}
		These equations describe the competition between inhomogeneous precession, driven by $\Delta_k$, and collective spin exchange, which couples the mode-resolved spins. They provide the classical mean-field dynamics associated with the underlying Heisenberg equations and establish the basis for analyzing exchange-assisted rephasing.

		It is useful to rewrite the coherent mean-field dynamics as a precession
		equation. Introducing
		\begin{equation}
			\mathbf u_k=(u_k^x,u_k^y,u_k^z),
			\qquad
			u_k^\pm=u_k^x\pm i u_k^y,
		\end{equation}
		the coherent dynamics can be written as
		\begin{equation}
			\label{eq:MeanFieldVectorForm}
			\left.
			\frac{d\mathbf u_k}{d\tau}
			\right|_{\rm coh}
			=
			\mathbf B_k^{\rm eff}\times \mathbf u_k,
		\end{equation}
		where
		\begin{equation}
			\label{eq:EffectiveExchangeField}
			\mathbf B_k^{\rm eff}
			=
			\Delta_k\hat{\mathbf z}
			+
			g_{\rm 0}
			\sum_{n\neq k}
			N_n U_{nk}^{\rm z}\mathbf u_n .
		\end{equation}
		The first term is the mode-dependent detuning responsible for coherent inhomogeneous phase mixing. The second is the self-consistent exchange field generated by the other occupied modes. Once the mode spins acquire different phases, the exchange field couples their directions and produces a collective torque that tends to reduce their relative phase dispersion. Exchange therefore acts as an autonomous internal feedback mechanism that can counteract coherent inhomogeneous phase mixing without measurement or externally conditioned control.

		The Ramsey contrast is the normalized collective transverse spin,
		\begin{equation}
			\label{eq:RamseyContrast}
			\mathcal{C}(\tau)
			=
			2\left|
			\sum_n p_n u_n^+(\tau)
			\right|,
		\end{equation}
		where the prefactor is chosen so that $\mathcal{C}(0)=1$ for an initially $x$-polarized spin coherent state.

		A rough criterion for interaction-assisted rephasing is that the collective exchange scale exceed the thermally induced detuning spread~\eqref{eq:DetuningSpread}, \(g_0N\gtrsim 12\delta_q/\eta^2\). Thus, the relevant inhomogeneity scale is set directly by the thermal broadening of the occupied modes
		\begin{equation}
			\label{eq:ExchangeDephasingParameter}
			\Lambda
			=
			\frac{g_0}{g_{0,c}}
			=
			\frac{g_0N\eta^2}{12\delta_q} = g_0N\tau_{\mathrm{mix}}.
		\end{equation}
		Thus, $\Lambda$ compares the collective exchange scale with the thermally induced inhomogeneous broadening and provides the natural parameter for comparing the dynamics at different temperatures; the system must compete with the generation of phase dispersion, not wait until the observable contrast has already collapsed.

		\begin{figure}[!t]
			\includegraphics[width = \columnwidth]{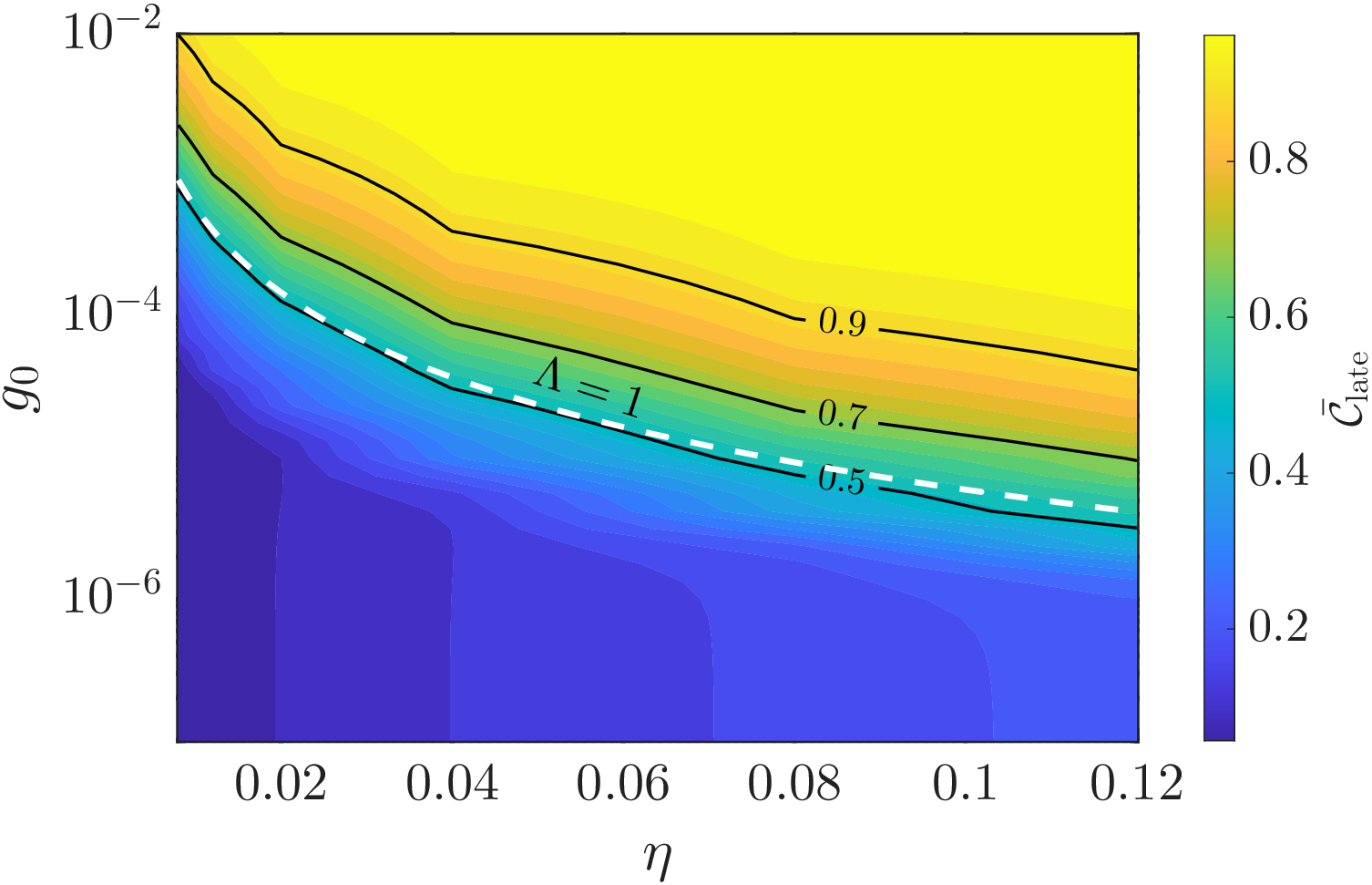}
			\caption{Late-time contrast landscape. The color scale shows $\bar{\mathcal{C} }_{\rm late}(\eta,g_{0})$, obtained by averaging $\mathcal{C}(s)$ over the late-time window. The dashed line marks $\Lambda =1$, corresponding to the point where the collective exchange scale becomes comparable to the analytic thermal inhomogeneous-broadening scale. 
			}\label{fig4}
		\end{figure}
        
		\subsection{Ramsey-contrast dynamics and coherence landscape}
			Figure~\ref{fig3} shows representative Ramsey-contrast dynamics for two values of the thermal parameter $\eta$, illustrating the crossover from inhomogeneous dephasing to interaction-assisted coherence recovery. The smaller value of $\eta$ corresponds to a broader thermal distribution over axial modes and therefore a larger spread of mode-dependent detunings, whereas larger $\eta$ narrows the occupied mode distribution and reduces the inhomogeneous broadening. 
			
			For weak coupling, $\Lambda\ll1$, the contrast collapses rapidly and remains small at late times. As $\Lambda$ approaches unity, partial recovery develops together with pronounced long-lived oscillations. For stronger coupling, $\Lambda>1$, the contrast remains substantially enhanced over the late-time observation window. Thus, increasing $\eta$ improves the recovered contrast at fixed $g_0$ by narrowing the thermal detuning spread. The observed crossover is therefore controlled not by temperature alone, but by the ratio of the collective exchange scale to the thermally induced inhomogeneous broadening.

			To characterize the long-time response after the initial dephasing transient, we define the late-time averaged contrast
			\begin{equation}
				\label{eq:lateTimeAverageContrast}
				\bar{\mathcal{C}}_{\rm late}(\eta,g_0) 	= \frac{1}{\tau_2-\tau_1} \int_{\tau_1}^{\tau_2} \mathcal{C}(\tau)\, d\tau ,
			\end{equation}
			where $\tau_1$ is chosen after the initial contrast collapse and $\tau_2$ defines the end of the averaging window.

			Figure~\ref{fig4} summarizes the long-time response over the interaction--temperature plane. The landscape shows a crossover from a low-contrast phase-mixed regime at weak coupling to a high-contrast regime at stronger coupling. The dashed line corresponding to $\Lambda=1$ tracks the onset of substantial contrast recovery, consistent with the criterion that the collective exchange scale becomes comparable to the thermally induced inhomogeneous broadening.
		
		\subsection{Scaling collapse and operational crossover threshold} 
		\begin{figure}[!t]
			\includegraphics[width = \columnwidth]{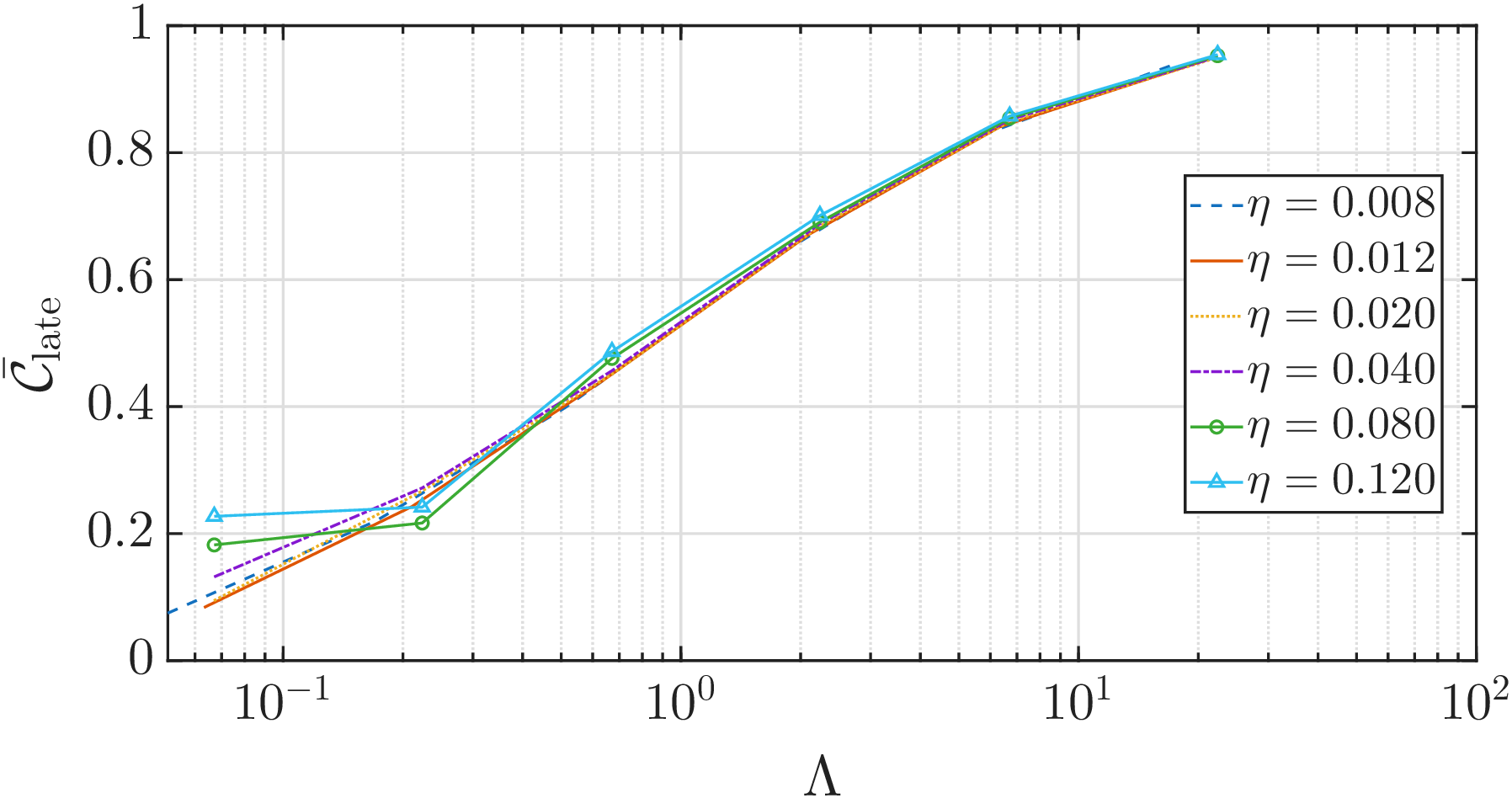}
			\caption{Scaling collapse of the late-time contrast. The late-time contrasts $\bar{\mathcal C}_{\rm late}$ versus exchange-to-inhomogeneity ratio $\Lambda$~\eqref{eq:ExchangeDephasingParameter} at different temperatures follow a common trend.
			}\label{fig5}
		\end{figure}
        
            The preceding analysis suggests that the dominant temperature dependence should be captured by the thermal broadening scale rather than by $\eta$ alone. Figure~\ref{fig5} tests this prediction by expressing the late-time averaged contrast as a function of the exchange-to-inhomogeneity ratio $\Lambda$. The results for different values of $\eta$ collapse onto a common trend, showing that the leading temperature dependence is captured by the analytic broadening scale. 
			
			The onset of substantial late-time contrast occurs near $\Lambda\sim1$, consistent with the criterion that the collective exchange scale becomes comparable to the thermally induced detuning spread. This collapse shows that the thermal dependence of the interaction-assisted recovery is largely organized by a single exchange-to-inhomogeneity ratio.

\begin{figure}[!t]
				\includegraphics[width = \columnwidth]{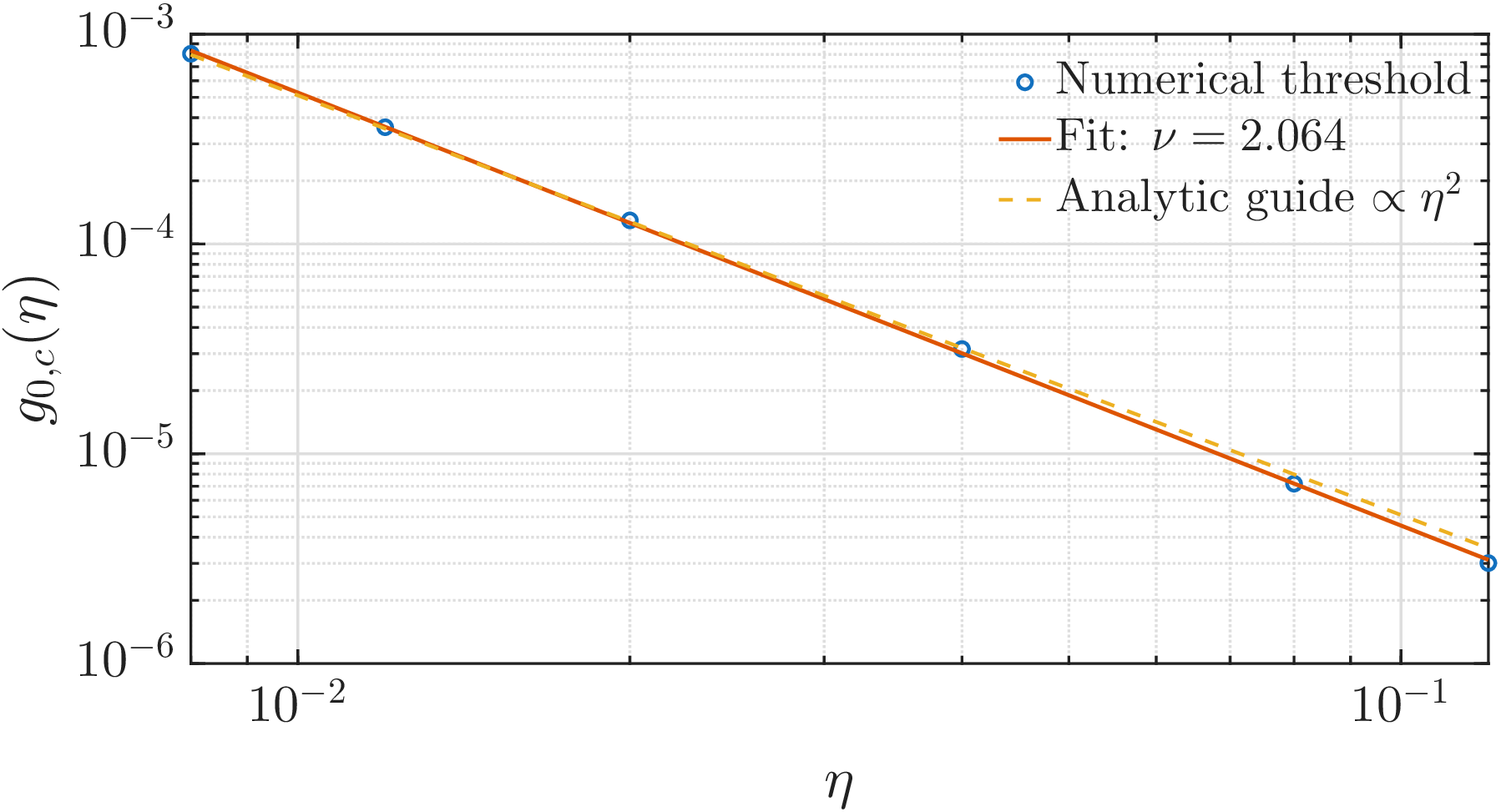}
				\caption{Operational crossover threshold. For each $\eta$, the threshold $g_{0,c}$ is defined by $\bar{\mathcal{C}}_{\rm late}(\eta,g_{0,c})=0.5$. The extracted threshold follows a power law $g_{0,c}\propto \eta^{-2.06}$, in close agreement with the analytic estimate $g_{0,c}\sim 	12\delta_q/(N\eta^2)$. The threshold therefore tracks the thermal inhomogeneous broadening scale.
				}\label{fig6}
			\end{figure}

			Finally, Fig.~\ref{fig6} provides an operational measure of the crossover. For each temperature, we define the threshold $g_{0,c}$ as the interaction strength for which the late-time averaged contrast reaches the reference value $\mathcal C_{\rm crit}=0.5$. The extracted threshold follows
			\begin{equation}
				g_{0,c}\propto\eta^{-2.06},
			\end{equation}
			in close agreement with the analytic estimate
			\begin{equation}
				g_{0,c}\sim\frac{12\delta_q}{N\eta^2}.
			\end{equation}
			Thus, the numerical crossover follows the same inverse-square temperature dependence as the analytically derived thermal inhomogeneous-broadening scale. The interaction strength required for substantial coherence recovery is therefore set by the width of the thermally occupied detuning distribution.

	\section{Robustness to Irreversible Dephasing}
	\label{sec:Robustness}
	
		The preceding analysis considered coherent inhomogeneous phase mixing, in which contrast loss arises from mode-dependent precession and can therefore be partially reversed by exchange dynamics. Realistic systems may also exhibit irreversible phase noise, such as residual fluctuations of the longitudinal detuning, which directly reduce transverse spin coherence. We incorporate this process through a pure-dephasing channel and examine how it modifies the mean-field dynamics. This allows us to distinguish exchange-assisted refocusing of coherent phase mixing from the irreversible loss of transverse coherence.
	
		\subsection{Pure dephasing}
	
			Residual fluctuations of the longitudinal detuning act as noise in the relative phase between the two internal states. Such noise damps the transverse spin components $S_n^\pm$, while leaving the longitudinal spin components unchanged. We model this effect using a pure-dephasing Lindblad term,
			\begin{equation}
				\frac{d\rho}{d\tau} = -i[\mathcal H,\rho] + \gamma \mathcal L[A]\rho ,
			\end{equation}
			where
			\begin{equation}
				\mathcal{L}[A]\rho = A\rho A^\dagger - \frac{1}{2} \left( A^\dagger A\rho + \rho A^\dagger A \right).
			\end{equation}
			The corresponding Heisenberg-picture evolution of an operator $\mathcal{O}$ is
			\begin{equation}
				\label{eq:AdjointLindbladEquation}
				\frac{d\mathcal{O}}{d\tau} =i[\mathcal H,\mathcal{O}] +\gamma \mathcal L^\dagger[A]\mathcal{O} ,
			\end{equation}
			with
			\begin{equation}
				\label{eq:AdjointLindblad}
				\mathcal{L}^\dagger[A]\mathcal{O} = A^\dagger \mathcal{O} A - \frac{1}{2} \left( A^\dagger A\mathcal{O} + \mathcal{O} A^\dagger A \right).
			\end{equation}
			For a Hermitian jump operator, \(A^\dagger=A\), this reduces to
			\begin{equation}
				\label{eq:AdjointDoubleCommutator}
				\mathcal{L}^\dagger[A]\mathcal{O} =	-\frac{1}{2}[A,[A,\mathcal{O}]] .
			\end{equation}
	
			At the mean-field level, we represent pure dephasing by a single transverse decay rate $\gamma_\perp$. For example, local pure dephasing of the form
			\begin{equation}
				\gamma_\phi \sum_{n,j}\mathcal{L}^\dagger[s_{n,j}^z]S^\pm_k = -\frac{\gamma_\phi}{2}S^\pm_k
			\end{equation}
			gives
			\begin{equation}
				\gamma_\perp=\frac{\gamma_\phi}{2}.
			\end{equation}
			For the mode-resolved variables \(u_k^+=\langle S_k^+\rangle/N_k\) and $u_k^z=\langle S_k^z\rangle/N_k$, the equations of motion become
			\begin{align}
				\label{eq:SplusEvolWithDephasing}
				\frac{du_k^+}{d\tau} &=	i\Delta_k u_k^+ + ig_{0} \sum_{n\neq k} N_n \tilde{U}_{nk}\left( 		u_k^+u_n^z-u_k^zu_n^+ \right) - \gamma_\perp u_k^+, \\
				\label{eq:SzEvolWithDephasing}
				\frac{du_k^z}{d\tau} &= \frac{ig_{0}}{2} \sum_{n\neq k} N_n \tilde{U}_{nk} \left(u_n^+u_k^- - u_n^-u_k^+ \right).
			\end{align}
	
			Equivalently, the same dynamics can be written in vector form as
			\begin{equation}
				\label{eq:MeanFieldVectorFormWithDephasing}
				\frac{d\mathbf u_k}{d\tau} = \mathbf B_k^{\rm eff}\times \mathbf u_k - \gamma_\perp \mathbf u_{k,\perp},
			\end{equation}
			where
			\begin{equation}
				\mathbf u_{k,\perp} = u_k^x\hat{\mathbf x} + u_k^y\hat{\mathbf y}.
			\end{equation}
			The first term of~\eqref{eq:MeanFieldVectorFormWithDephasing} describes coherent precession in the self-consistent exchange field, while the second produces irreversible transverse damping. Thus, exchange can reorganize and partially refocus coherent mode-dependent phase mixing, but pure dephasing introduces an additional loss channel that limits the recovered Ramsey contrast.
			
				\begin{figure}[!t]
					\includegraphics[width = \columnwidth]{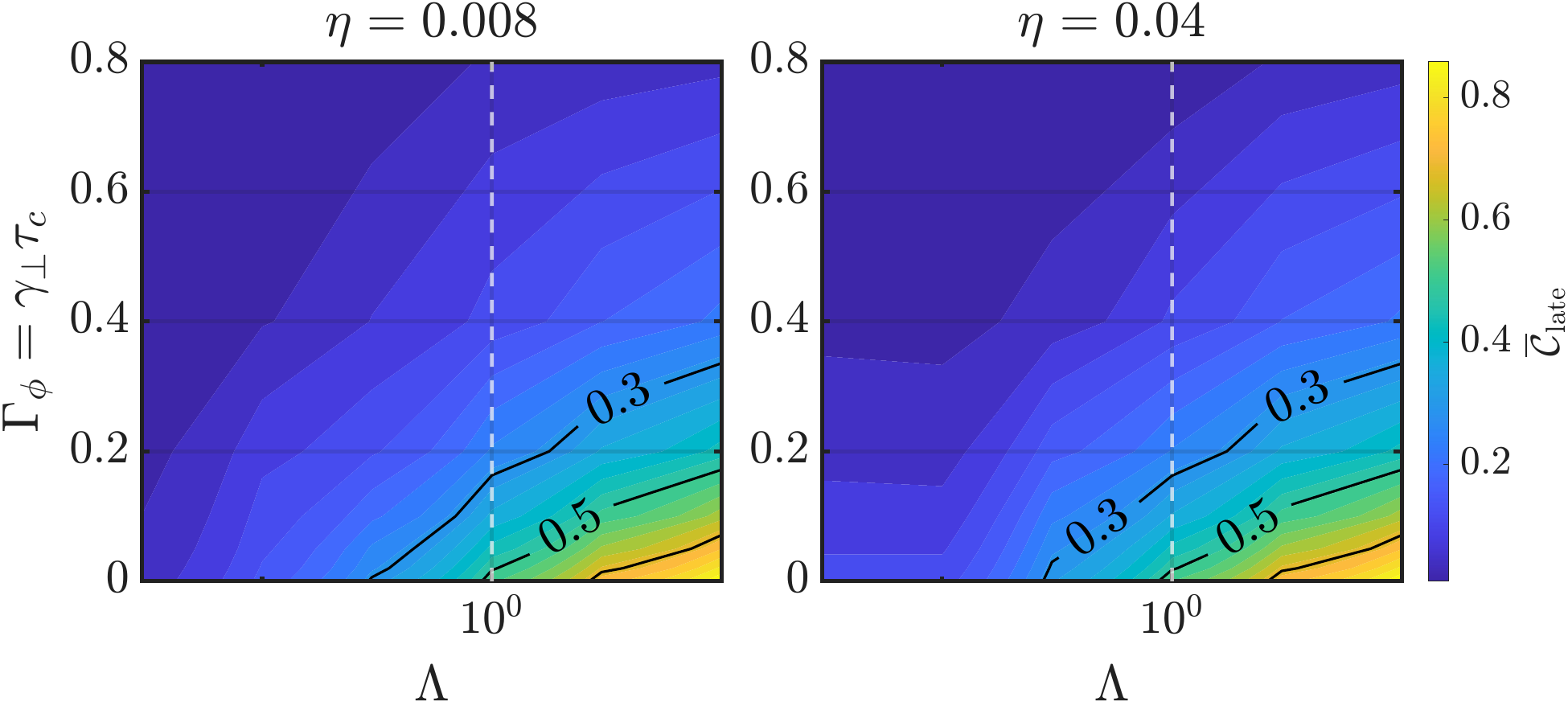}
					\caption{Robustness of exchange-assisted rephasing to pure dephasing. The color scale shows the late-time Ramsey contrast $\bar{\mathcal C}_{\rm late}$ as a function of the scaled exchange strength $\Lambda$ and the accumulated dephasing strength  $\Gamma_\phi=\gamma_\perp\tau_c$, for two representative temperatures. The vertical dashed line marks $\Lambda=1$. 
					}\label{fig7}
				\end{figure}

			\subsection{Robustness to pure dephasing}
				
				For comparison across temperatures, we characterize dephasing by the accumulated dimensionless decay over the collapse time,
				\begin{equation}
					\Gamma_\phi=\gamma_\perp\tau_c, \qquad \tau_c=\frac{\eta}{12\delta_q}.
				\end{equation}
				Thus, in the absence of interactions and inhomogeneous precession, the transverse coherence would acquire the envelope
				\begin{equation}
					e^{-\gamma_\perp\tau} = e^{-\Gamma_\phi s}, \qquad 	s=\frac{\tau}{\tau_c}.
				\end{equation}
				The parameter \(\Gamma_\phi\) measures the accumulated irreversible transverse decay on the scaled time interval used to characterize the late-time contrast $\bar{\mathcal C}_{\rm late}$~\eqref{eq:lateTimeAverageContrast}.

				Figure~\ref{fig7} shows the late-time contrast as a function of the scaled exchange strength $\Lambda$~\eqref{eq:ExchangeDephasingParameter} and the accumulated dephasing strength $\Gamma_\phi$, for two representative temperatures. The high-contrast region occurs for large $\Lambda$ and small $\Gamma_\phi$. Increasing $\Lambda$ enhances the late-time contrast, whereas increasing $\Gamma_\phi$ suppresses it. The contour lines therefore bend upward with increasing $\Lambda$: stronger exchange is required to maintain the same late-time contrast in the presence of stronger irreversible dephasing. Thus, maintaining a given level of recovered contrast in the presence of stronger irreversible dephasing requires stronger exchange.

				The vertical dashed line marks $\Lambda=1$, the crossover scale identified from the coherent inhomogeneous-broadening problem. For weak dephasing, this line remains close to the onset of exchange-assisted recovery. Pure dephasing therefore does not eliminate this crossover identified in Section~\ref{sec:AssistedRephasing}, but introduces a second, independent limitation on the attainable contrast. Even when $\Lambda\gtrsim1$, the recovered contrast decreases as $\Gamma_\phi$ increases. The similarity of the two temperature panels further indicates that the robustness is governed primarily by the scaled variables $\Lambda$ and $\Gamma_\phi$, rather than by temperature alone.

					\begin{figure}[!t]
						\includegraphics[width = \columnwidth]{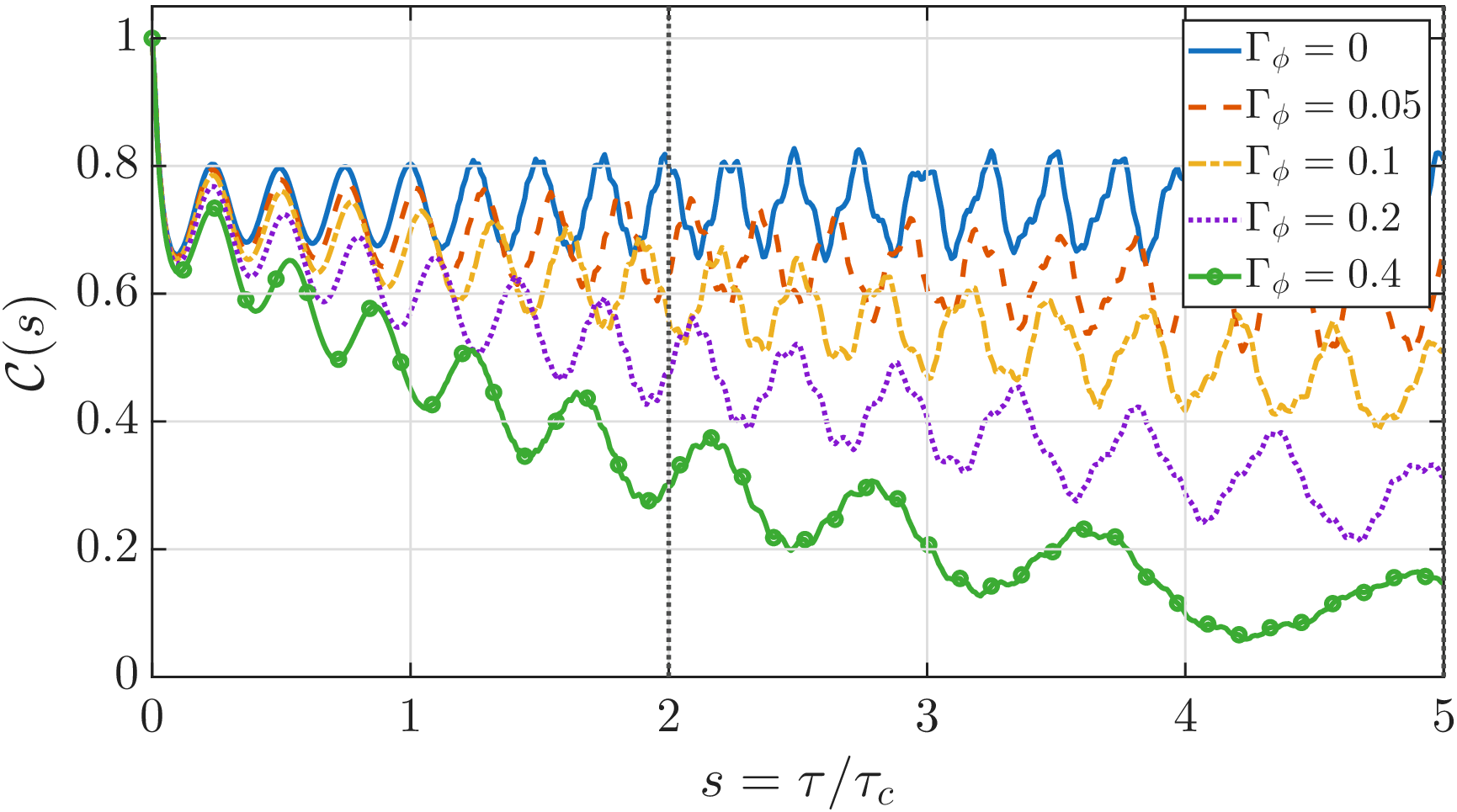}
						\caption{Time-domain signature of pure dephasing in the exchange-rephased regime. The Ramsey contrast $\mathcal C(s)$ is shown for $\eta=0.04$ and $\Lambda=3$, for several values of $\Gamma_\phi=\gamma_\perp\tau_c$. The vertical dotted line at $s=2$ marks the beginning of the late-time averaging window.
						}\label{fig8}
					\end{figure}
				
				The corresponding time-domain behavior is shown in Fig.~\ref{fig8} for dimensionless temperature $\eta=0.04$ and $\Lambda = 3$, where exchange produces a strongly recovered  contrast in the absence of dephasing. As $\Gamma_\phi$ is increased, the oscillatory exchange-rephased signal remains visible, but its envelope and late-time mean value are progressively reduced. Thus, pure dephasing does not simply shift the coherent exchange dynamics; it continuously removes transverse coherence from the collectively rephased state.

				These results clarify the role of interactions in the presence of irreversible noise. Exchange can counteract coherent inhomogeneous phase mixing by reducing the relative phase dispersion among the mode spins and reorganizing their dynamics into a collective response. It cannot, however, undo local irreversible phase noise. The exchange-assisted recovery is therefore robust to weak pure dephasing, but the maximum recoverable contrast is limited by the accumulated transverse decay measured by $\Gamma_\phi$. Thus exchange reorganizes the coherent inhomogeneous dynamics into a partially rephased collective state, whose  observable Ramsey contrast is ultimately limited by the accumulated transverse decay.

	\section{Summary and Conclusions}\label{sec:Summary}

		We have developed a weak-axis-resolved Hamiltonian theory of interaction-assisted spin coherence in a thermally occupied multimode trapped atomic ensemble. Starting from contact interactions and retaining the secular forward-scattering channels, we obtain a mode-resolved spin Hamiltonian combining mode-dependent longitudinal detunings with overlap-dependent exchange couplings. The axial mode structure is retained explicitly, while the transverse degrees of freedom are incorporated through thermal averaging. This construction provides the microscopic origin of the self-consistent exchange field that governs the mean-field rephasing dynamics.

        In the absence of interactions, Ramsey-contrast loss arises from coherent phase mixing among thermally occupied modes. Near the magic magnetic field, the residual weak-axis shift is quartic in position, producing a nonlinear distribution of mode-dependent precession frequencies. In the many-mode continuum regime, the resulting contrast develops an inverse-square-root stationary-phase envelope rather than an exponential decay, establishing the noninteracting benchmark and the relevant thermal inhomogeneous-broadening scale.

        At the mean-field level, the exchange generates a restoring torque that reduces the relative phase dispersion of the mode spins and restores collective transverse coherence. The dynamics is organized by the competition between the collective exchange scale and the thermally induced detuning spread. When expressed in terms of the corresponding exchange-to-inhomogeneity ratio, results at different temperatures collapse onto a common trend, while the operational crossover coupling follows the predicted \(g_{0,c}\propto\eta^{-2}\) scaling, with a fitted exponent of approximately \(2.06\). Substantial interaction-assisted coherence therefore emerges when the collective exchange scale becomes comparable to the thermal broadening of the occupied modes.

        Irreversible pure dephasing provides a distinct limitation. Weak dephasing preserves the exchange-assisted response over a finite observation window while progressively reducing the attainable contrast. Exchange can refocus deterministic mode-dependent phase dispersion, but it cannot reconstruct coherence destroyed by stochastic phase noise. Taken together, these results connect microscopic trapped-gas interactions to mode-resolved self-rephasing and establish a simple physical criterion for interaction-assisted coherence: exchange can sustain collective spin coherence against coherent inhomogeneous broadening when its collective scale competes with the detuning spread, whereas irreversible decoherence remains an independent limitation.

	\begin{acknowledgments}
		The authors acknowledge support from William \& Mary start-up funds. Computational resources were provided by the SciClone high-performance computing facility at William \& Mary.
	\end{acknowledgments}

	\appendix
	
	\section{Inhomogeneous Detuning \label{sec:InhomogeneousDetuning}}
	Consider a trapped atom containing two internal clock states, denoted by $\lvert\uparrow\rangle$ and $\lvert\downarrow\rangle$, generally experience different Zeeman shifts in the presence of a magnetic field. Consequently, the transition frequency between the two states becomes position dependent when the magnetic field varies across the trapped cloud~\cite{harber2002,deutsch2010}. In second quantization, the single-particle Hamiltonian for the two-level system is
	\begin{equation}
		\label{eq:SingleParticleHamil}
		\hat{H}^\Delta = \int\,d\mathbf{r}\, \hat{\boldsymbol{\Psi}}^\dagger(\mathbf{r})\left[\frac{\mathbf{p}^2}{2M} + V_\mathrm{trap}(\mathbf r) + \frac{\hbar}{2}\Delta(\mathbf r)\sigma^z\right] \hat{\boldsymbol{\Psi}}(\mathbf{r}),
	\end{equation}
	where $\hat{\boldsymbol\Psi}(\mathbf r) = \left( \hat{\Psi}_\uparrow(\mathbf r), \hat{\Psi}_\downarrow(\mathbf r) \right)^T$, $\hat{\Psi}_\sigma(\mathbf r)$ annihilates an atom in internal state $\sigma \left(= \uparrow,\,\downarrow\right)$ at position $\mathbf r$, $\sigma^z$ is the Pauli $z$ operator acting in the two-level internal-state space, and $\hbar\Delta (\mathbf{r})$ is the local clock shift. Let $\phi_n(\mathbf{r})$ be the harmonic trap basis with the corresponding eigenvalue $E_n$. The field operators $\hat{\Psi}_\sigma$ are expanded in the harmonic trap basis $\phi_n(\mathbf{r})$, 
	\begin{equation}
		\label{eq:OperatorExpansion}
		\hat{\Psi}_\sigma(\mathbf r) = 	\sum_n \phi_n(\mathbf r)\, \hat{c}_{\sigma n}.
	\end{equation}
	Upon substituting~\eqref{eq:OperatorExpansion} in~\eqref{eq:SingleParticleHamil}, the off-diagonal terms of the single particle Hamiltonian acquire phases, $e^{i(E_n-E_m)t/\hbar}$, in the interaction picture with respect to the trap Hamiltonian. If the differential shift is weak on the scale of the trap level spacings, these rapidly oscillating terms may be neglected within a secular, or random-phase, approximation. Equivalently, the differential shift is replaced by its diagonal matrix elements $\tilde{\Delta}_n\equiv \tilde{\Delta}_{nn}$, giving 
	\begin{equation}
		\label{eq:Dephasing}
		\hat{H}^\Delta = \hbar \sum_n \tilde{\Delta}_n	S^z_n
	\end{equation}
	where $\tilde{\Delta}_n$ is as defined in the main text,~\eqref{eq:DeltaNCompact}.
	
	\section{Many-Body Dynamics\label{sec:ManyBody}}
	Also atoms in the trap experience contact two-body interactions~\cite{vankempen2002,chin2010},
	\begin{equation}
		\label{eq:ContactInteraction}
		\hat{H}^{\mathrm{int}} = \frac{1}{2} \sum_{\sigma,\sigma' = \uparrow,\downarrow} g_{\sigma\sigma'} \int d^3r \, \hat{\Psi}^\dagger_\sigma(\mathbf r) \hat{\Psi}^\dagger_{\sigma'}(\mathbf r)  	\hat{\Psi}_{\sigma'}(\mathbf r)
		\hat{\Psi}_\sigma(\mathbf r).
	\end{equation}
	The contact interactions \eqref{eq:ContactInteraction} are characterized by $	g_{\sigma\sigma'} = 4\pi\hbar^2 a_{\sigma\sigma'}/M$ that determine the collision strength between particles in different spins states, $\sigma,\, \sigma'$, where $a_{\sigma\sigma'}$ is the $s$-wave scattering length for collisions between atoms in internal states $\sigma$ and $\sigma'$. Upon substituting~\eqref{eq:OperatorExpansion} in~\eqref{eq:ContactInteraction} each quartic term acquires the phase $e^{i(E_n+E_m-E_p-E_q)t/\hbar}$ in the interaction picture with respect to the trap Hamiltonian. Terms for which the incoming and outgoing motional energies are nonresonant therefore dephase rapidly and are neglected within a secular, or random-phase, approximation. In the stronger forward-scattering approximation, we retain only those collision processes that preserve the set of occupied motional modes, namely the direct channel $(q,p)=(n,m)$ and the exchange channel $(q,p)=(m,n)$~\cite{oktel2002,williams2002,mcguirk2002}. 
	
	\subsection{Direct Term}
	The direct term contains only density-density interactions. It corresponds to forward scattering processes in which the motional modes of the atoms are unchanged. Taking $p=m$ and $q=n$, we obtain
	\begin{equation}
		\label{eq:DirectTerm}
		H^D = \sum_{\substack{n,m\\ n<m}} U_{nm} \sum_{\sigma,\sigma'} g_{\sigma\sigma'} \hat{c}^\dagger_{n\sigma}\hat{c}_{n\sigma} \hat{c}^\dagger_{m\sigma'}\hat{c}_{m\sigma'},
	\end{equation}
	where
	\begin{equation}
		\label{eq:Overlap}
		U_{nm} = \int d^3r\, \lvert\phi_n(\mathbf r)\rvert^2 \lvert\phi_m(\mathbf r)\rvert^2.
	\end{equation}

	Using the number operator $N_k = \hat{c}^\dagger_{\uparrow k} \hat{c}_{\uparrow k } + \hat{c}^\dagger_{\downarrow k} \hat{c}_{\downarrow k} $, and the mode-resolved spin operators
	\begin{equation}
		\label{eq:ModeLadderOperators}
		\begin{split}			
			S^z_k & = \frac{\hat{c}^\dagger_{\uparrow k} \hat{c}_{\uparrow k } - \hat{c}^\dagger_{\downarrow k} \hat{c}_{\downarrow k}}{2}\\
			S^+_k & = \hat{c}^\dagger_{k\uparrow}\hat{c}_{k\downarrow}, \\
			S^-_k & = \hat{c}^\dagger_{k\downarrow}\hat{c}_{k\uparrow}.
		\end{split}
	\end{equation} 
	the direct interaction Hamiltonian becomes
	\begin{equation}
		\label{eq:HDirect}
		\begin{split}
			H^D = \sum_{\substack{n,m\\ n<m}} U_{nm} \Bigg[ & \frac{ g_{\uparrow\uparrow} + g_{\downarrow\downarrow} +
				2g_{\uparrow\downarrow} }{4} \hat{N}_n\hat{N}_m  \\
			& + 	\frac{ g_{\uparrow\uparrow} - g_{\downarrow\downarrow}}{2} \left( \hat{N}_m S^z_n + \hat{N}_n S^z_m \right) \\
			& + \left(g_{\uparrow\uparrow} + g_{\downarrow\downarrow} - 2g_{\uparrow\downarrow} \right) S^z_nS^z_m \Bigg],
		\end{split}
	\end{equation}
	where $g_{\downarrow\uparrow} = g_{\uparrow\downarrow}$.

	\subsection{Exchange term}
	The exchange term corresponds to forward-scattering processes in which the two 	atoms exchange their internal states between motional modes. Taking $p=n$ and $q=m$, we obtain
	\begin{equation}
		\label{eq:ExchangeInteraction}
		H^{\rm ex} = \sum_{\substack{n,m\\ n<m}} U_{nm} \sum_{\sigma,\sigma'} g_{\sigma\sigma'} \hat{c}^\dagger_{n\sigma} 		\hat{c}_{n\sigma'} 	\hat{c}^\dagger_{m\sigma'} \hat{c}_{m\sigma}.
	\end{equation}
	
	Using the number operators  $N_n = \hat{c}^\dagger_{\uparrow n} \hat{c}_{\uparrow n } + \hat{c}^\dagger_{\downarrow n} \hat{c}_{\downarrow n} $ and the mode-resolved spin operator \eqref{eq:ModeLadderOperators} the exchange Hamiltonian becomes
	\begin{equation}
		\label{eq:ExHamil}
		\begin{split}
			H^{\rm ex} = &\sum_{\substack{n,m\\ n<m}} U_{nm} \Bigg[ \frac{ g_{\uparrow\uparrow} + g_{\downarrow\downarrow} }{4}
			\hat{N}_n\hat{N}_m\\ 
			& + \frac{ g_{\uparrow\uparrow} - g_{\downarrow\downarrow}}{2} \left(\hat{N}_m S^z_n + \hat{N}_n S^z_m \right) + \left( g_{\uparrow\uparrow} + g_{\downarrow\downarrow} \right) S^z_nS^z_m \\
			& + g_{\uparrow\downarrow} \left( S^+_nS^-_m + S^-_nS^+_m \right) \Bigg].
		\end{split}
	\end{equation}
	
	\subsection{Inter-Mode Forward-Scattering Hamiltonian}
	Thus the two-body interaction Hamiltonian $H^D+ H^\mathrm{ex}$ becomes
	\begin{equation}
		\label{eq:InterModeHamil}
		\begin{split}
			H^\mathrm{FSA}_\mathrm{inter} = &\sum_{\substack{n,m\\n<m}}U_{nm}\Bigg[\frac{g_{\uparrow\uparrow} + g_{\downarrow\downarrow} + g_{\uparrow\downarrow}}{2}\hat{N}_n\hat{N}_m \\
			& +\left(g_{\uparrow\uparrow} - g_{\downarrow\downarrow}\right)\left(\hat{N}_mS^z_n + \hat{N}_nS^z_m\right) \\
			& + 2\left(g_{\uparrow\uparrow}+ g_{\downarrow\downarrow} - g_{\uparrow\downarrow}\right)S^z_nS^z_m + g_{\uparrow\downarrow}\left(S^+_nS^-_m + S^-_nS^+_m\right)
			\Bigg]
		\end{split}
	\end{equation}
	
	The sum over $n<m$ describes interactions between atoms occupying distinct motional modes. Same-mode interaction terms with $n=m$ are not included in this inter-mode sum. 
	

	\subsection{Same-Mode Forward-Scattering Hamiltonian}	
	These same-mode interactions is given by 
	\begin{equation}
		\label{eq:SameModeHamil}
		H^{\mathrm{FSA}}_\mathrm{same} = \frac{1}{2}\sum_{n} U_{nn}\sum_{\sigma\sigma'}g_{\sigma\sigma'}\hat{c}^\dagger_{n\sigma}\hat{c}^\dagger_{n\sigma'}\hat{c}_{n\sigma'}\hat{c}_{n\sigma}.
	\end{equation}
	More explicitly in terms of number operators
	\begin{equation}
		\label{eq:SameModeMain}
		\begin{split}
			H^{\mathrm{FSA}}_\mathrm{same} =&  \frac{1}{2}\sum_{n} U_{nn}\Bigg[ g_{\uparrow\uparrow}\hat{N}_{n\uparrow}\left(\hat{N}_{n\uparrow} - 1\right)\\ 
			& + g_{\downarrow\downarrow}\hat{N}_{n\downarrow}\left(\hat{N}_{n\downarrow} - 1\right) + 2g_{\uparrow\downarrow}\hat{N}_{n\uparrow}\hat{N}_{n\downarrow}\Bigg]
		\end{split}
	\end{equation}
	This form makes it clear that no one atom interacts with itself.
	
	Using the number operator $N_k = \hat{c}^\dagger_{\uparrow k} \hat{c}_{\uparrow k } + \hat{c}^\dagger_{\downarrow k} \hat{c}_{\downarrow k} $ and the mode-resolved spin operators~\eqref{eq:ModeLadderOperators}, one writes \eqref{eq:SameModeMain} in terms of spin operators
	\begin{equation}
		\label{eq:SameModeSpinForm}
		\begin{split}
			H^{\mathrm{FSA}}_{\mathrm{same}} = \frac{1}{2} \sum_n U_{nn} \Bigg\{
			& \frac{ g{\uparrow\uparrow} + g_{\downarrow\downarrow} + 2g_{\uparrow\downarrow} }{4} \hat N_n^2 -
			\frac{ g_{\uparrow\uparrow} + g_{\downarrow\downarrow} }{2} \hat N_n \\
			& + \left( g_{\uparrow\uparrow} - g_{\downarrow\downarrow} \right) \left( \hat N_n-1 \right) S_n^z \\
			& + \left( g_{\uparrow\uparrow} + g_{\downarrow\downarrow} - 2g_{\uparrow\downarrow} \right) \left( S_n^z \right)^2 \Bigg\}.
		\end{split}
	\end{equation}
	
	\subsection{Interaction Hamiltonian}
	Hence the interaction Hamiltonian is 
	\begin{equation}
		H^\mathrm{int} = H^\mathrm{FSA}_\mathrm{inter} + H^\mathrm{FSA}_\mathrm{same}.
	\end{equation} 
	
	However, in the spin-independent limit, $g_{\uparrow \uparrow} = g_{\downarrow\downarrow} = g_{\uparrow\downarrow} \equiv g$, the spin-dependent terms of $H^\mathrm{FSA}_\mathrm{same}$ vanish, and the same-mode Hamiltonian $H^\mathrm{FSA}_\mathrm{same}$~\eqref{eq:SameModeSpinForm} reduces to $$ H^\mathrm{FSA}_\mathrm{same} = \frac{g}{2}\sum_n U_{nn} \hat{N}_n\left(\hat{N}_n - 1\right).$$ Thus, for isotropic interactions and fixed mode occupations, the same-mode term is only a density-dependent energy shift. This term depends only on the mode occupations and is therefore spin independent. For fixed occupations \(\{N_n\}\), it produces only an overall energy shift and does not affect the mode-resolved spin dynamics. It can be dropped from the spin Hamiltonian used to describe exchange-assisted rephasing. If mode-changing collisions or density dynamics are retained, this term contributes to the occupation-dependent motional interaction energy, but those processes are outside the forward-scattering spin model considered here.
	
	On the other hand, within the spin-independent limit, $g_{\uparrow \uparrow} = g_{\downarrow\downarrow} = g_{\uparrow\downarrow} \equiv g$, the inter-mode forward-scattering Hamiltonian $H^\mathrm{FSA}_\mathrm{inter}$~\eqref{eq:InterModeHamil} separates into a density-density shift and an isotropic exchange interaction,
	\begin{equation}
		\label{eq:InterHamilReduced}
		\begin{split}
			H^\mathrm{FSA}_\mathrm{inter} = &\sum_{\substack{n,m\\n<m}}U_{nm}\Bigg[\frac{3g}{2}\hat{N}_n\hat{N}_m \\
			& + 2g\left(S^z_nS^z_m + \frac{1}{2}\left(S^+_nS^-_m + S^-_nS^+_m\right)\right)
			\Bigg]
		\end{split}
	\end{equation}
	For fixed mode occupations, the first term is a spin-independent density shift and does not affect the mode-resolved spin dynamics. Dropping this constant shift gives the effective spin-exchange Hamiltonian~\eqref{eq:EffectiveSpinHamiltonian}, used in the main text
	\begin{equation}
		\label{eq:InterSpinExchangeHamiltonian}
		H^\mathrm{int}  = 2g \sum_{n<m} U_{nm} \left[ S_n^zS_m^z  + \frac{1}{2} \left( S_n^+S_m^- + S_n^-S_m^+ \right) \right].
	\end{equation}
	This isotropic mode-spin exchange structure is the microscopic counterpart of the exchange-driven collective dynamics appearing in kinetic and energy-class descriptions of spin self-rephasing~\cite{piechon2009,deutsch2010,maineult2012,bonnin2019}.

	   \bibliography{ReferenceFile}

\end{document}